\documentclass[12pt]{iopart}

\expandafter\let\csname equation*\endcsname\relax
\expandafter\let\csname endequation*\endcsname\relax
\usepackage{amsmath}
\usepackage{graphicx}
\begin{document}

\title[Chaotic transients and hysteresis in an $\alpha^2$ dynamo model]{Chaotic transients and hysteresis in an $\alpha^2$ dynamo model}

\author{Dalton N. Oliveira}
\address{Institute of Aeronautical Technology - ITA, 12228-900, S\~ao Jos\'e dos Campos, SP, Brazil}
\ead{dalton@ita.br}
\author{Erico L. Rempel}
\address{Institute of Aeronautical Technology - ITA, 12228-900, S\~ao Jos\'e dos Campos, SP, Brazil}
\ead{rempel@ita.br}
\author{Roman Chertovskih}
\address{Research Center for Systems and Technologies (SYSTEC), Faculdade de Engenharia da Universidade do Porto, Rua Dr. Roberto Frias, s/n 4200-465, Porto, Portugal }
\ead{roman@fe.up.pt}
\author{Bidya B. Karak}
\address{Department of Physics, Indian Institute of Technology (Banaras Hindu University), Varanasi 221005, India}
\ead{karak.phy@iitbhu.ac.in}

\date{\today}

\begin{abstract}
The presence of chaotic transients in a nonlinear dynamo is investigated through numerical simulations of the 3D magnetohydrodynamic equations. By using the kinetic helicity of the flow as a control parameter, a hysteretic blowout bifurcation is conjectured to be responsible for the transition to dynamo, leading to a sudden increase in the magnetic energy of the attractor. This high-energy hydromagnetic attractor is suddenly destroyed in a boundary crisis when the helicity is decreased. Both the blowout bifurcation and the boundary crisis generate long chaotic transients that are due, respectively, to a chaotic saddle and a relative chaotic attractor.  
\end{abstract}


\section{Introduction}
\label{introduction}

Chaotic transients are a common phenomenon in fluids and plasmas, being usually associated with decaying turbulence, where an initially erratic fluid converges to a laminar state. A typical example is a pipe flow, where turbulent puffs can last for a long time, but eventually disappear if the pipe is sufficiently long and the Reynolds number is below a certain threshold \cite{Hof08}. 
Chaotic transients are known to be due to the presence of nonattracting chaotic sets in the phase space \cite{Kantz87,Hsu88}.
In two or more dimensional phase spaces, these nonattracting chaotic sets have a stable and an unstable manifold, which are 
main directions of attraction and repulsion; the nonattracting chaotic set lies in the intersection of both manifolds and is, then, dully called a chaotic saddle \cite{Nusse89}. In spatially extended systems,
where the phase space is infinite-dimensional, chaotic saddles may be responsible for transient temporal chaos \cite{Rempel03}
or transient spatiotemporal chaos \cite{Rempel07}.
In space and astrophysical plasmas, chaotic transients related to chaotic saddles have been observed in numerical simulations of Alfv\'en waves \cite{Chian07}, magnetohydrodynamic (MHD) dynamo \cite{Rea09} and accretion disks \cite{Rempel2010}.  
The goal of the present paper is to study the appearance of chaotic transients in an MHD simulation of transition to dynamo, a crucial topic in the study of the origin and evolution of astrophysical magnetic fields.

Many astrophysical systems -- such as planets, stars, and galaxies -- show variable magnetic fields.
In some cases, the field even shows coherent structures and polarity reversals. The characteristic spatial length-scale of such magnetic fields is comparable to the size of the system and thus this is technically called the large-scale field, 
in contrast to the small-scale one which has a characteristic length-scale smaller or equal to the characteristic scale of the driving flow \cite{BS05}. As an example, the solar and stellar magnetic cycles are a manifestation of the large-scale field.
 

A large-scale dynamo is responsible for the generation and maintenance of large-scale magnetic fields and cycles in all astrophysical bodies mentioned above. The plasma in such bodies usually displays differential rotation, where the angular velocity is not uniform but varies according to the latitude. Depending on the relative importance of the differential rotation, which gives rise to the so-called $\Omega$--effect, and the helical convection, which gives rise to the so-called $\alpha$--effect  \cite{KR80}, 
the large-scale dynamo is characterized as $\alpha^2$ or $\alpha$$\Omega$ dynamo. 
When the differential rotation is significant, the dynamo is of $\alpha$$\Omega$ type. 
On the other hand, when the differential rotation is negligible, 
the dynamo is said to be of $\alpha^2$ type. The solar and galactic dynamos are thus of $\alpha$$\Omega$ type. However, there can be some objects
in which the rotation is very uniform, like the Earth, and the dynamo is of $\alpha^2$ type. In rapidly rotating stars, 
the differential rotation is highly quenched, while the $\alpha$ effect remains strong and thus the dynamo is expected to be of $\alpha^2$ type \cite{CK06}.

With the evolution of the astrophysical bodies over the age, the dynamo efficiency may change. 
For example, the stellar rotation decreases with the age (primarily due to magnetic braking \cite{Skumanich72}) 
and this reduces the efficiency of the dynamo \cite{HN87}.
When the rotation becomes sufficiently slow, the dynamo number ($D$) goes below a critical value ($D_c$)
and the large-scale dynamo ceases to operate \cite{KR80}. There are indications that the solar dynamo is probably slightly above this critical value \cite{R84, Met16, KN17}.
However, the dynamo can still persist when $D< D_c$.
Both the mean-field dynamo model \cite{KO10} and magnetohydrodynamic (MHD) simulations \cite{KKB15} reveal that near the onset of the large-scale dynamo, two stable
states can coexist and attract different initial conditions. The magnetic field vanishes when started with a weak initial field but
the magnetic energy remains high (the strong-field branch) when started with a strong field.
Thus, the dynamo displays a hysteresis near the dynamo onset. 

The motivation of the present study is to explore the origin of chaotic transients nearby the onset of an $\alpha^2$ dynamo in the presence of hysteresis. 
For this, we consider a simple 3D MHD model of isothermal compressible fluid
which is driven by a helical forcing function. The advantage of including an external driver is that we can vary the net helicity in our model to explore the dynamo transition. In section \ref{model}, the $\alpha^2$ dynamo model is described; section \ref{results} shows the main results of the paper and the conclusions are given in section \ref{conclusion}.

\section{THE MODEL}
\label{model}

We adopt the model of $\alpha^{2}$ dynamo employed in Refs.~\cite{Brandenburg2001, Rempel2013}. The fluid is assumed to be isothermal and compressible, with constant sound speed $c_{s}$, constant dynamical viscosity $\mu$, constant magnetic diffusivity $\eta$ and constant magnetic permeability $\mu_{0}$. The governing equations are: 

\begin{gather}
\partial_{t} \ \textrm{ln}\rho + \textit{\textbf{u}} \cdot \nabla \ \textrm{ln}\rho + \nabla \cdot \textit{\textbf{u}} = 0, \label{eqn:continuity}\\
\partial_{t}\textit{\textbf{u}} + \textit{\textbf{u}} \cdot \nabla \textit{\textbf{u}} = -\nabla{p}/ \rho + \textit{\textbf{J}} \times \textit{\textbf{B}} / \rho + (\mu / \rho) \left( \nabla^{2} \textit{\textbf{u}} + \nabla \nabla \cdot \textit{\textbf{u}}/3\right) + \textit{\textbf{f}},  \label{eqn:momentum}\\
\partial_{t}\textit{\textbf{A}} = \textit{\textbf{u}} \times \textit{\textbf{B}} -\eta \mu_{0}\textit{\textbf{J}}, \label{eqn:induction}
\end{gather}
 where $\rho$ is the density, \textit{\textbf{u}} is the fluid velocity, $\textit{\textbf{A}}$ is the magnetic vector potential, $\textit{\textbf{J}} = \nabla  \times \textit{\textbf{B}}/ \mu_{0}$ is the current density, \textit{p} is the pressure, $\textit{\textbf{f}}$ is an external forcing function and $\nabla p / \rho = \textit{c}^{2}_{s} \nabla \ \textrm{ln}\rho $ where $ \textit{c}^{2}_{s} = \gamma p /\rho $ is assumed to be constant. The magnetic induction equation (\ref{eqn:induction}) is written for the vector potential $\textit{\textbf{A}}$ to ensure a solenoidal magnetic field, since $\nabla \cdot \textit{\textbf{B}} = \nabla \cdot (\nabla \times \textit{\textbf{A}}) = 0$. The logarithmic density is also adopted in Eqs. (\ref{eqn:continuity}) and (\ref{eqn:momentum}) for numerical reasons, since it varies spatially much less than density.
The domain is a box with dimensions $L_{x} = L_{y} = L_{z} = 2 \pi$ and periodic boundary conditions in all three directions for all variables. We adopt non-dimensional units with $ k_{1} = \textit{c}_{s} = \rho_{0} = \mu_{0} = 1$, where $\rho = \langle \rho \rangle$ is the spatial average of $\rho$ and $k_{1}$ is the smallest wave number in the simulation box. Time is measured in units of $ \left( \textit{c}_{s}k_{1}\right)^{-1} $, space is measured in units of $k^{-1}_{1}$, $\textit{\textbf{u}}$ in units of $ \textit{c}_{s}$,  $\textit{\textbf{B}}$ in units of $\sqrt{\mu_{0} \rho_{0}} \textit{c}_{s}$, $\rho$ in units of $\rho_{0}$ and the magnetic diffusivity $\eta$ is in units of $\textit{c}_{s}/k_{1}$.

Equations (\ref{eqn:continuity})-(\ref{eqn:induction}) are solved with the PENCIL CODE\footnote{http://pencil-code.nordita.org/}, a thoroughly tested MHD solver frequently employed in Astrophysical works (see, e.g., \cite{Axel2019} and the journal issue dedicated to the physics and algorithms of the PENCIL CODE). The code adopts sixth-order finite differences in space and third-order variable step Runge-Kutta in time. The initial conditions are  ln $\rho$ = 0 and  \textit{\textbf{u}} = 0.  The initial magnetic vector potential is modeled by noise with Gaussian distribution with zero mean and standard deviation equal to $10^{-3}$. Kinetic energy is injected in the system by a forcing function \textit{\textbf{f}}, which is defined by \cite{Brandenburg2001,Axel2001b}

\begin{equation}
\textit{\textbf{f}}(\textit{\textbf{x}}, t) = \textrm{Re} \left\lbrace \textit{N} \textit{\textbf{f}}_{\textit{\textbf{k}}(t)} \textrm{exp}\left[ i \textit{\textbf{k}}(t) \cdot \textit{\textbf{x}} + i \phi(t)\right] \right\rbrace,
\label{eqn:forcing}
\end{equation}
where $\textit{\textbf{k}}(t) = (k_{x}, k_{y}, k_{z})$ is a time dependent wave vector, $\textit{\textbf{x}} = (x,y,z)$ is position, and $\phi(t)$, with  $\vert \phi \vert < \pi$, is a random phase. Here, $\textit{N} = \textit{f}_{0}\textit{c}_{s} \left( \textit{k}\textit{c}_{s} / \delta t \right)^{1/2}$, where $\textit{f}_{0}$ is a nondimensional factor, $ \textit{k} = \vert \textit{\textbf{k}} \vert $, and $\delta t$ is the integration time step. We choose the forcing wave number $k$ around $\textit{k}_{f} = 5$ and every time step, a vector $\textit{\textbf{k}}(t)$ with $4.5 < k < 5.5$ is randomly selected from a set of 350 previously generated vectors with
the given wave number. The operator $\textit{\textbf{f}}_{\textit{\textbf{k}}}$ is given by

\begin{equation} 
\textit{\textbf{f}}_{\textit{\textbf{k}}} = \frac{\textit{\textbf{k}} \times (\textit{\textbf{k}} \times  \hat{\textit{\textbf{e}}}) - i\sigma k (\textit{\textbf{k}} \times \hat{\textit{\textbf{e}}})}{\sqrt{1+\sigma^{2}} \textit{k}^{2} \sqrt{1-(\textit{\textbf{k}} \cdot \hat{\textit{\textbf{e}}})^{2}/ \textit{k}^{2}}}, \label{forcing}
\end{equation}
where $\hat{\textit{\textbf{e}}}$  is an arbitrary unit vector needed in order to generate a vector $\textit{\textbf{k}} \times \hat{\textit{\textbf{e}}}$ which is perpendicular to $\textit{\textbf{k}}$. Note that for $\sigma = 1$, $\vert \textit{\textbf{f}}_{\textit{\textbf{k}}}\vert^{2} = 1$ and the kinetic helicity density of this forcing function satisfies
\begin{equation}
\textit{\textbf{f}} \cdot \nabla \times  \textit{\textbf{f}} = \vert \textit{\textbf{k}}\vert \textit{\textbf{f}}^{\hspace{0.075cm} 2}> 0  
\end{equation}
at each point in space. For $\sigma=0$, the forcing function is nonhelical, so $\sigma$ is a measure of the kinetic helicity of the forcing. This rather complex forcing function has been employed in several previous works for being able to generate turbulent statistics even for moderate Reynolds numbers \cite{haugen} and for us it is interesting because it allows us to choose the level of helicity added to the flow, an important element for the generation of large-scale dynamos.

The Reynolds number and magnetic Reynolds number are defined, respectively, as 
\begin{equation}
R_{e} = \frac{\textit{u}_{rms}}{\nu k_{f}}, \quad R_{m}=\frac{\textit{u}_{rms}}{\eta k_{f}},
\end{equation}
where $\nu = \mu/ \rho$ is the kinematic viscosity, $\textit{u}_{rms} \equiv\left \langle \textit{\textbf{u}}^{2} \right \rangle ^{1/2}$ is the root-mean-square (r.m.s) velocity and spatial average is denoted by $ \left \langle \cdot   \right \rangle$.

\section{Results}
\label{results}

\begin{figure}[th!]\begin{center}
\includegraphics[width=\columnwidth]{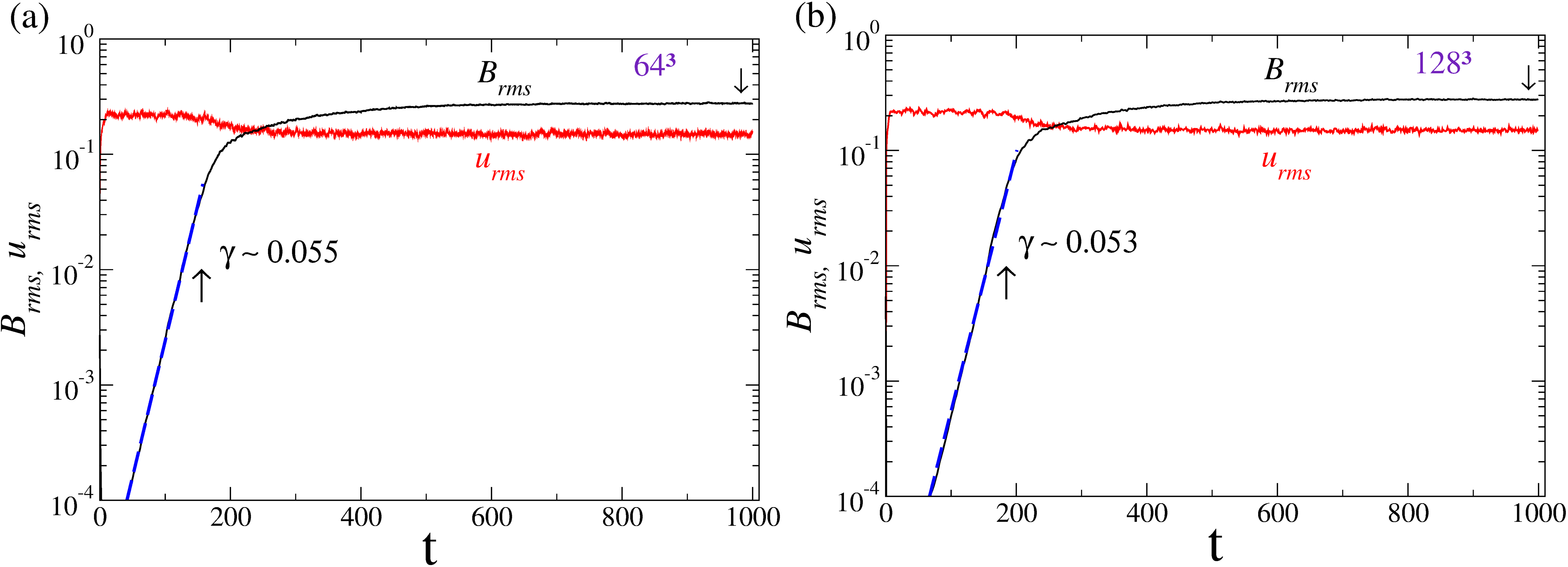}
\end{center}\caption[]{
 Comparison of numerical simulations with different grid resolutions. (a) Time series of $\textit{B}_{rms}$ (black line) and $\textit{u}_{rms}$ (red line)  of MHD dynamo simulations in log--linear scale for $\sigma = 1$ using $64^3$ grid points; the kinematic phase has a growth rate of $\gamma \approx 0.055$. (b) The same as (a), but for $128^3$ grid points; the growth rate is $\gamma \sim 0.053$.  }\label{time-series1}\end{figure}

First we must choose the grid resolution for the numerical simulation of Eqs. (\ref{eqn:continuity})-(\ref{eqn:induction}). Figure~\ref{time-series1} shows a comparison of the time series of the root mean square of the magnetic ($B_{rms}$, black line) and velocity ($u_{rms}$, red line) fields in log--linear axes for simulations using $64^3$ (a)
and $128^3$ (b) grid points. The parameter values are $f_0=0.07$, $\nu=\eta=0.005$ and $\sigma=1$. Initially, the seed magnetic field is too weak to impact the velocity field, but the velocity field has a strong impact on the induction equation (\ref{eqn:induction}), causing an exponential growth of the magnetic energy during the initial (kinematic) phase of the dynamo. The growth rate $\gamma$ in this phase can be found as the slope of a fitted line in the log-linear plot, and it is approximately 0.05 for both resolutions. As the magnetic flux grows, it starts to strongly affect the velocity field through the Lorentz force in the momentum equation (\ref{eqn:momentum}), causing a rapid decrease in the kinematic energy around  $t=200$ for both resolutions. Eventually, the mean magnetic and kinetic energies reach a saturated value which is approximately the same for both resolutions, for which we have $R_e = R_m \approx 20$. Based on that, and considering the large amount of long time series that must be computed in this work, we adopted the lower resolution of $64^3$ points in the following sections.

Next, we set $f_0=0.07$, $\nu=\eta=0.005$ and vary $\sigma$ as a control parameter.
Using $\sigma$ as the control parameter is a natural choice, since it is known that the presence of kinetic helicity 
in the flow can be favorable for
magnetic field generation \cite{GFP}, although it is not strictly needed for either small- or large-scale dynamo
to operate (see \cite{RCZ,ACZ} and references therein). 

\begin{figure}[th!]
\centering
\includegraphics[width=0.7\columnwidth]{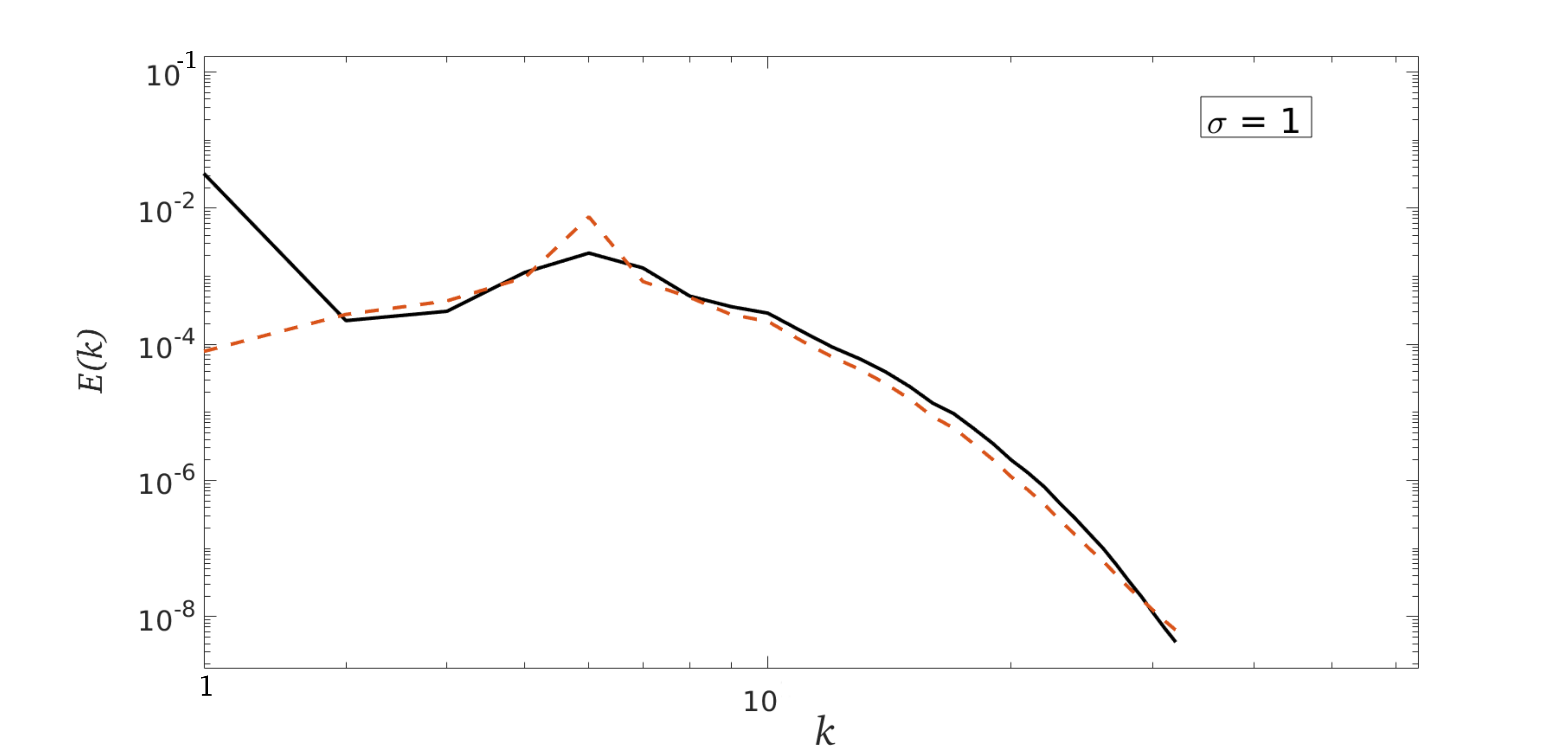}
\caption{\label{fig:powerspectrum1} Spectra of kinetic  (red dashed line) and magnetic (black solid line) energies  at times  $t = 1000$ for $\sigma=1$.}
\end{figure}

\begin{figure}[th!]\begin{center}
\includegraphics[width=\columnwidth]{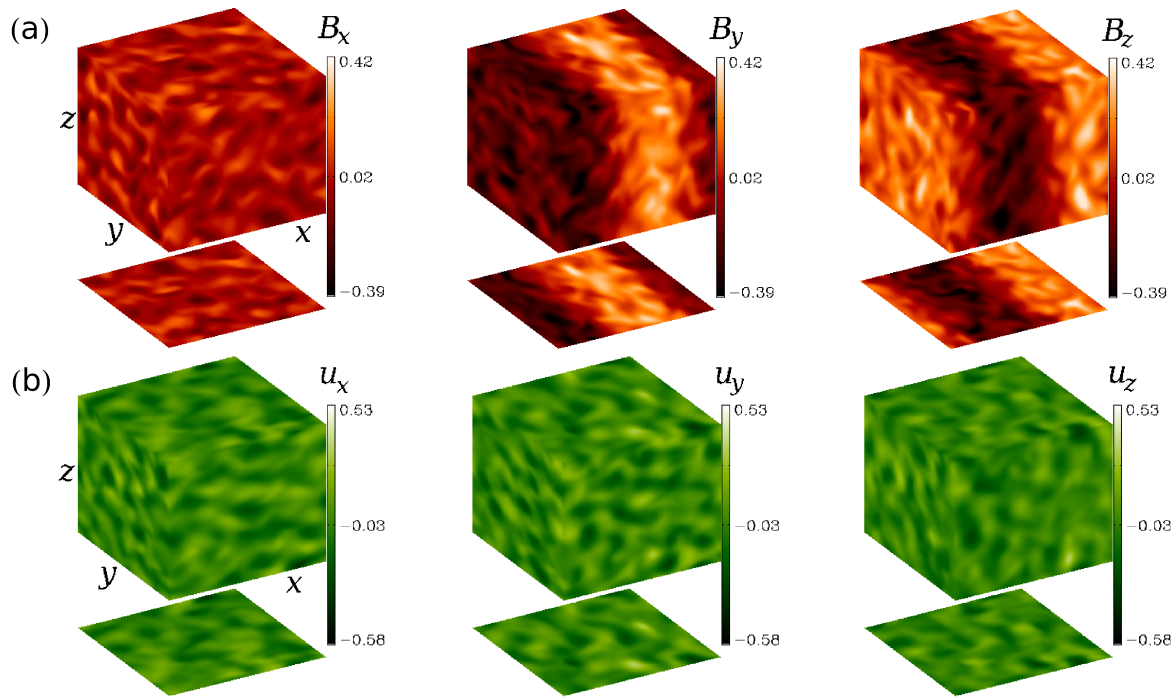}
\end{center}\caption[]{
Intensity plot of magnetic field (a) and velocity field (b) components at $t=1000$ for $\sigma=1.$}\label{fig:sigma09-1}\end{figure}

Since our forcing function has $k$ around $k_f=5$, kinetic energy is injected at this scale in the flow, inducing the production of a series of eddies with that wave number in the physical space. When $\sigma=1$, helicity is maximum in the flow, causing an inverse energy transfer from $k=5$ to $k=1$ that has been related to the $\alpha$--effect in \cite{Brandenburg2001}. 
This can be observed by plotting the one-dimensional power spectra, as in Fig. \ref{fig:powerspectrum1}, where the kinetic (red dashed line) and magnetic (black solid line) spectra are computed as the integrated energy along spherical shells in the $\textit{\textbf{k}}=(k_x, k_y, k_z)$ space for $t=1000$, well inside the nonlinearly saturated dynamo regime. It is clear that the kinetic spectrum is peaked at $k=5$ and the magnetic at $k=1$. This causes the appearance of large scale magnetic structures that are illustrated in the upper panel of Fig. \ref{fig:sigma09-1}. Note that $B_y$ and $B_z$ display a large scale oscillation in addition to the small scale fluctuations. The lower panel shows the velocity field components, whose largest structures are at the $k_f$ scale.   

\subsection{Bifurcation diagram}
\label{results2}

\begin{figure}[th!]\begin{center}
\includegraphics[width=0.7\columnwidth]{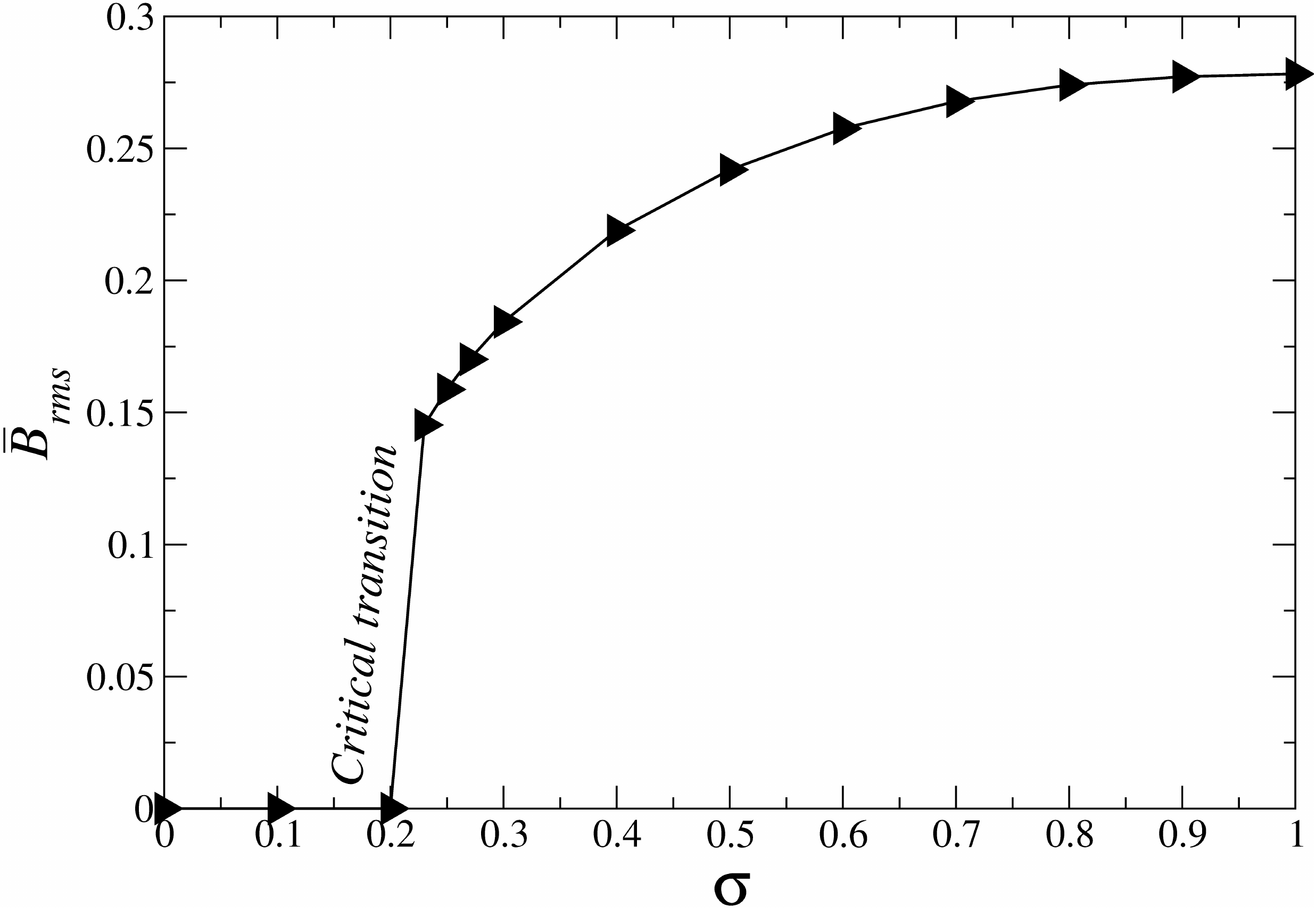}
\end{center}\caption[]{
Bifurcation diagram showing the time-averaged $B_{rms}$ as a function of $\sigma$.}\label{diagram-bifurcation}\end{figure}

The onset of dynamo action is shown in Figure  \ref{diagram-bifurcation}, which represents the bifurcation diagram as a function of $\sigma$ for the time-averaged root mean square magnetic field $\bar{B}_{rms}$. For each value of $\sigma$, a weak seed magnetic field is used as initial condition, the initial transient  is discarded and then, time averages of ${B}_{rms}$ are plotted. The magnetic energy of the attractor is zero when $\sigma$ is below 0.21, implying that there is no dynamo action and the attracting state is purely hydrodynamic. For $\sigma > 0.21$ dynamo action takes place and there is a sudden jump in the $B_{rms}$ of the attractor. The average magnetic energy of this {\em hydromagnetic attractor} keeps growing from then on, up to $\bar{B}_{rms}\approx 0.275$ for $\sigma=1$. Note that at the critical transition, the hydrodynamic attractor loses stability and small amplitude magnetic perturbations are sufficient to drive the system toward the hydromagnetic attractor.

\begin{figure}[th!]\begin{center}
\includegraphics[width=1\columnwidth]{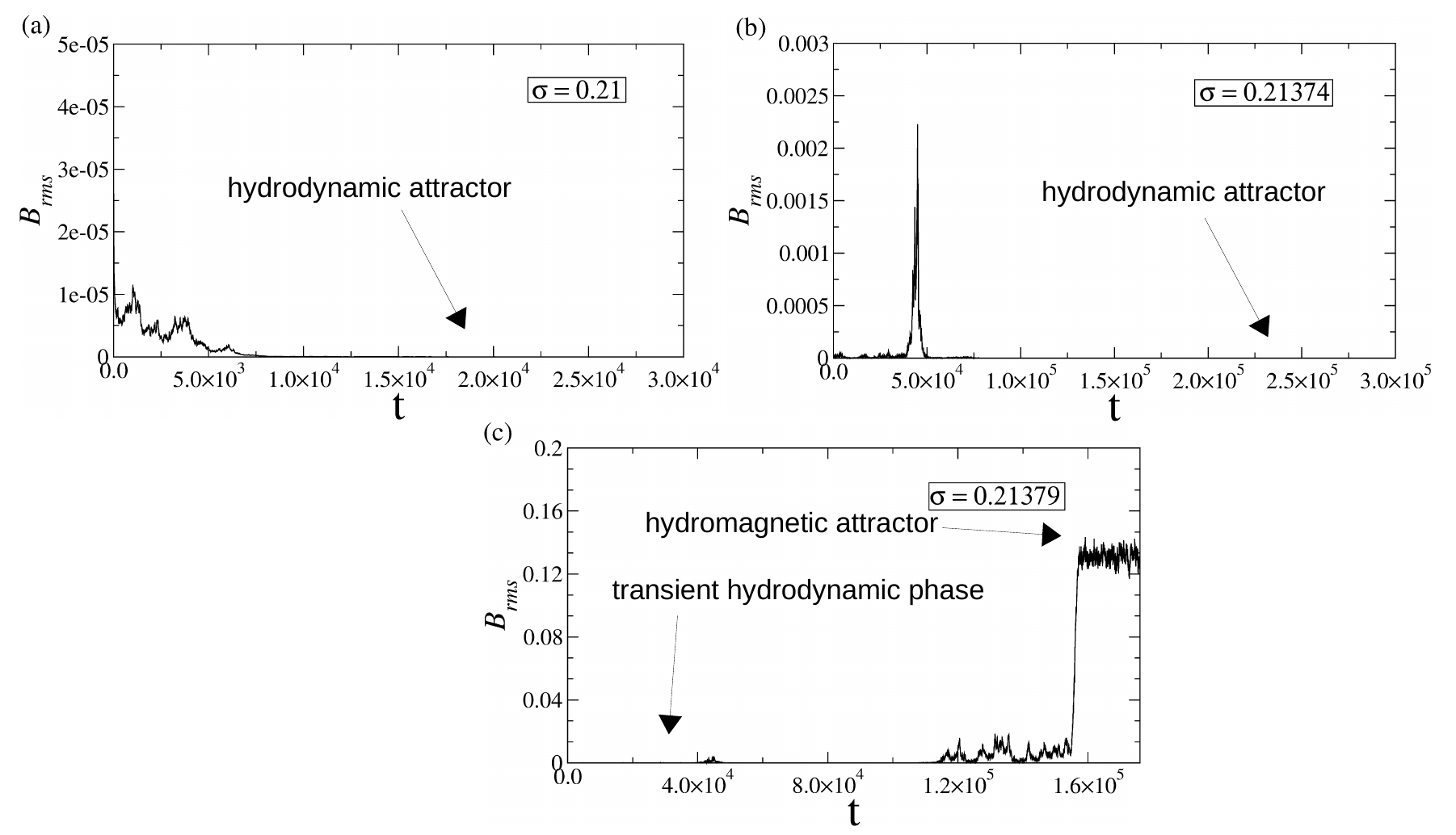}
\end{center}\caption[]{
 Time series of $B_{rms}$ nearby the critical transition to dynamo action shown in Fig. \ref{diagram-bifurcation}, for a random seed magnetic field. Before the transition (a and b), showing decay to a purely hydrodynamic attractor, and right after the transition (c), showing a long transient hydrodynamic phase before reaching to the hydromagnetic attractor.
 }\label{serie-021}\end{figure}

In previous works \cite{Sweet01}, transition to dynamo in MHD simulations in a periodic box with the helical ABC-forcing was shown to be due to a nonhysteretic {\em blowout bifurcation}. In this type of bifurcation, the dynamical system has a smooth invariant manifold within which lies a chaotic attractor for parameter values less than a critical value. As the parameter value is increased, a blowout bifurcation takes place, in which the manifold loses its attracting property and the chaotic set on it ceases to be an attractor. Right after the transition, solutions display {\em on--off intermittency}, i.e., they spend long times very near the manifold, then are ``blown out" of it in fast bursts where they move far from the manifold. After each burst, trajectories go back to the vicinity of the manifold and the process repeats intermittently \cite{Ott94}. We tried in vain to find intermittent bursts in the transition to dynamo using the forcing function in the form (\ref{eqn:forcing}). Figures \ref{serie-021}(a)-(b) show that just before the transition, small magnetic perturbations decay and the solutions converge to the purely hydrodynamic ($\textit{\textbf{B}}=0$) manifold. Thus, this manifold is attracting and there is a chaotic hydrodynamic attractor in it (velocity field fluctuations are always chaotic in our work). Very near the transition, some magnetic bursts may occur as in Fig. \ref{serie-021}(b) for $\sigma=0.21374$, but they have a tiny amplitude and, eventually, the hydrodynamic  manifold attracts the solution and no more bursts are observed. For $\sigma=0.21379$, right after the transition, the solution stays near the manifold for a long time before suddenly jumping toward a chaotic attractor with high magnetic energy, the hydromagnetic attractor. This shows that the hydrodynamic manifold has lost transversal stability, since initial conditions with   
$\textit{\textbf{B}}=0$ stay on the manifold for all parameter values, but even small nonzero values of $\textit{\textbf{B}}$ (i.e., perturbations that are transversal to the hydrodynamic manifold) are able to expel trajectories away from the manifold and toward the hydromagnetic attractor. This means that the previous hydrodynamical attractor has also lost its transversal stability and what is left is a transient chaotic hydrodynamic phase. It could be said that the  hydrodynamic chaotic attractor has become a chaotic saddle, but since this chaotic set still attracts all initial conditions on the hydrodynamic manifold, which itself has become unstable, we refer to this hydrodynamic chaotic set as a {\em relative chaotic attractor}, adopting a nomenclature introduced by Skufca et al. \cite{Skufca06}. Note that the stable manifold of the relative attractor in question is not a fractal structure, but the whole hydrodynamic subspace defined by $\textit{\textbf{B}}=0$. The reason for the absence of high-amplitude intermittent bursts in our dynamo transition is explained in the next section.

\subsection{Hysteresis and chaotic transients}

\begin{figure}[th!]\begin{center}
\includegraphics[width=0.7\columnwidth]{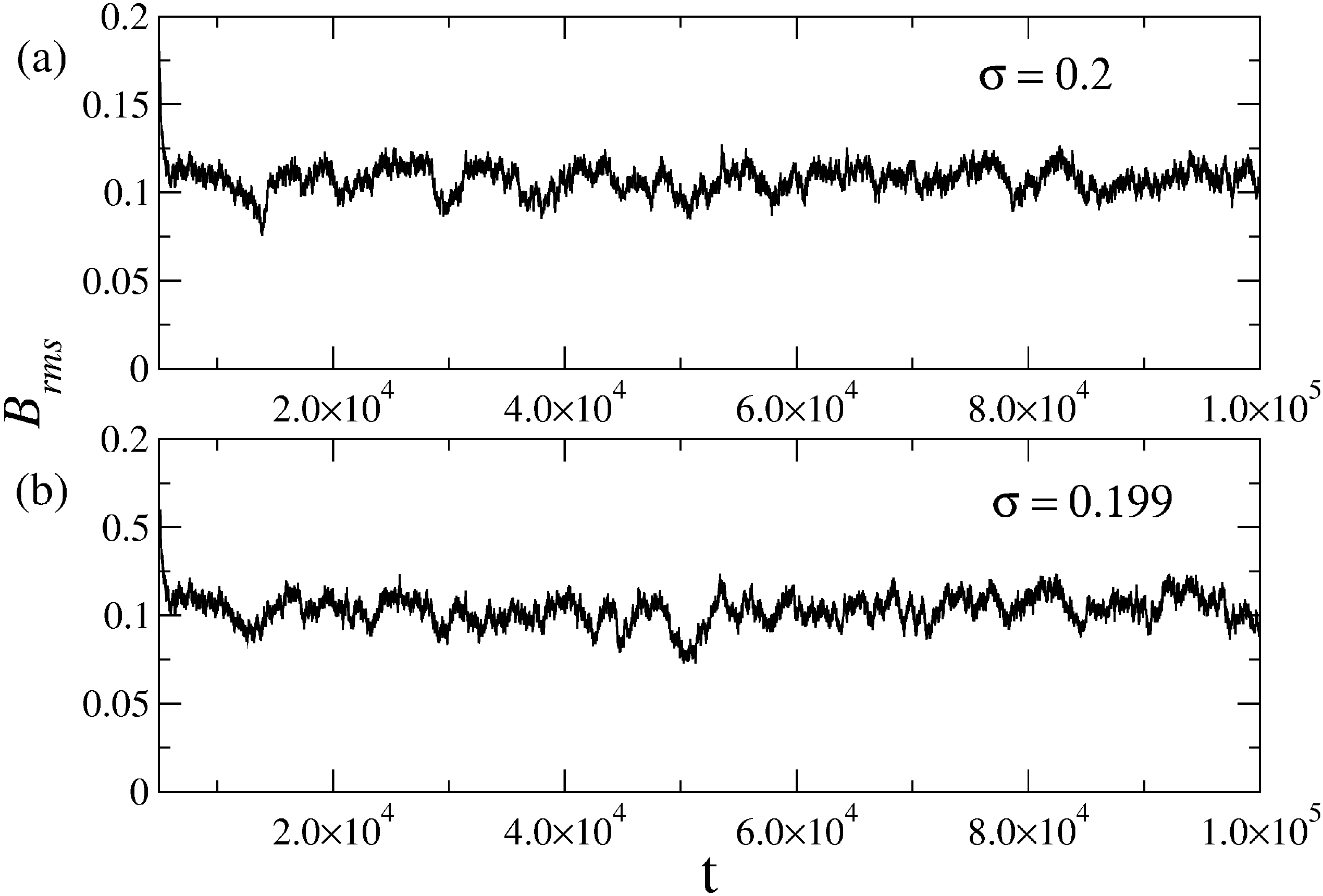}
\end{center}\caption[]{
 Time series of $B_{rms}$ for two different values of $\sigma$ smaller than the critical transition to dynamo shown in Fig. \ref{diagram-bifurcation}. The simulations started from a strong initial magnetic field obtained at $\sigma = 0.3$. Then the control parameter was reduced to (a) $\sigma = 0.2$ and (b) $\sigma = 0.199$. The solution does not decay toward the hydrodynamic attractor with null magnetic energy,
indicating hysteresis.}\label{histeresetime}\end{figure}

Motivated by Karak et al. \cite{KKB15}, we search for hysteresis in this dynamo system. Although their work has an
imposed uniform large-scale shear flow that is absent in our simulations, we still managed to find a hysteresis
in our $\alpha^2$--dynamo model. Recall from section \ref{results2} that the transition to the hydromagnetic attractor takes place at $\sigma = \sigma_c \approx 0.21379$, where random seed magnetic fields are amplified; for $\sigma<\sigma_c$, the seed fields decay to zero, as the solutions approach a hydrodynamic attractor. However, if one takes as initial condition a state with a high energy magnetic field, it may not decay to the hydrodynamic attractor, as shown in Fig. \ref{histeresetime}. Here, the initial condition is a state taken from the hydromagnetic attractor at $\sigma=0.3$; when the control parameter is reduced to $\sigma=0.2$ (Fig. \ref{histeresetime}(a)) and $\sigma=0.199$ (Fig. \ref{histeresetime}(b)), the solutions remain in the hydromagnetic attractor, with the magnetic energy still following the upper branch in Fig. \ref{diagram-bifurcation}. That is a signature of a hysteresis, as the saturation of the magnetic field amplitude depends on the previous history of the control parameter.

\label{multistability}

\begin{figure}[th!]\begin{center}
\includegraphics[width=0.7\columnwidth]{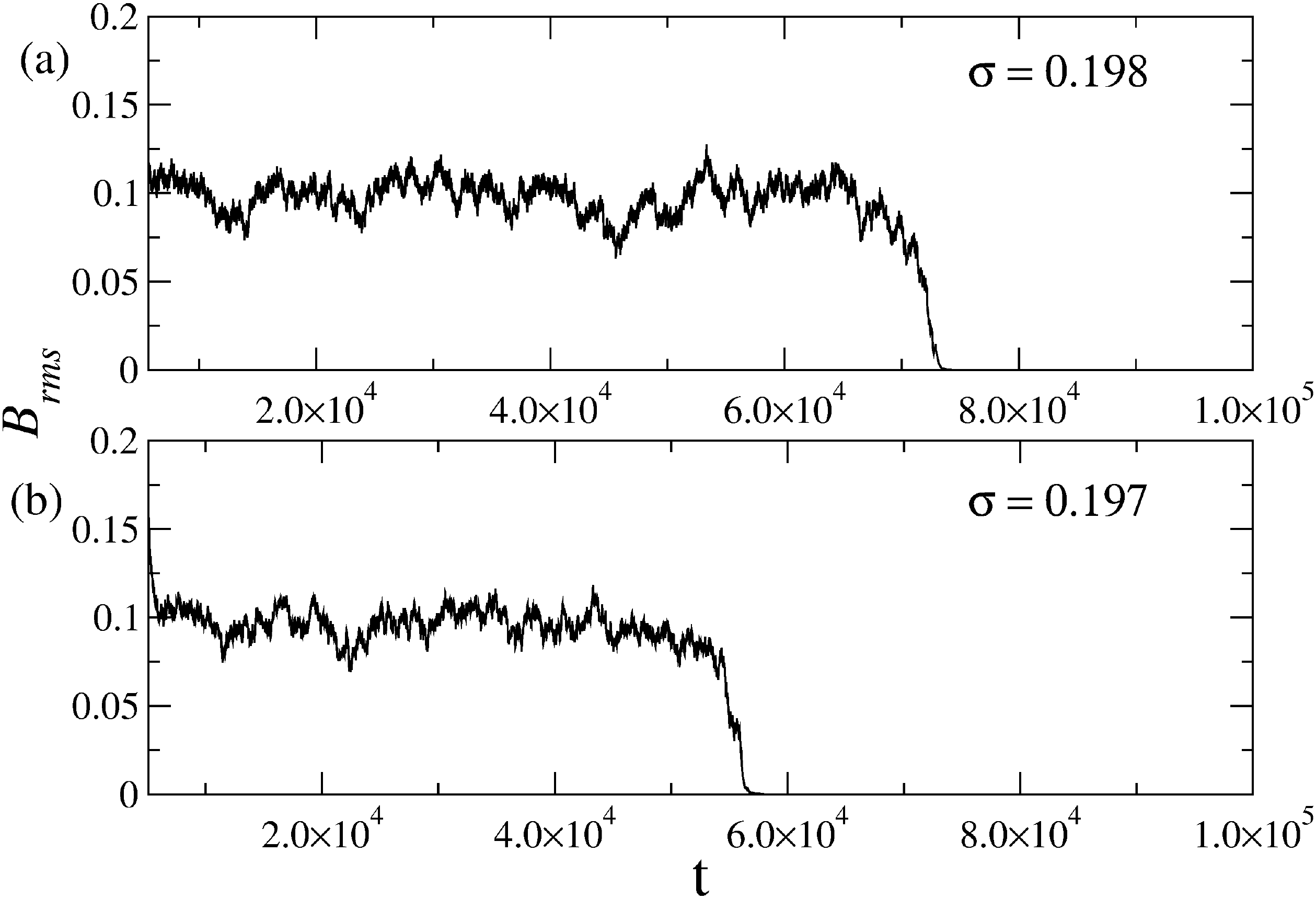}
\end{center}\caption[]{
Time series of $B_{rms}$ exhibiting hydromagnetic chaotic transients for (a) $\sigma = 0.198$ and (b) $\sigma = 0.197$.
 }\label{fig:transient-0198-0197}\end{figure}

\begin{figure}[th!]\begin{center}
\includegraphics[width=0.7\columnwidth]{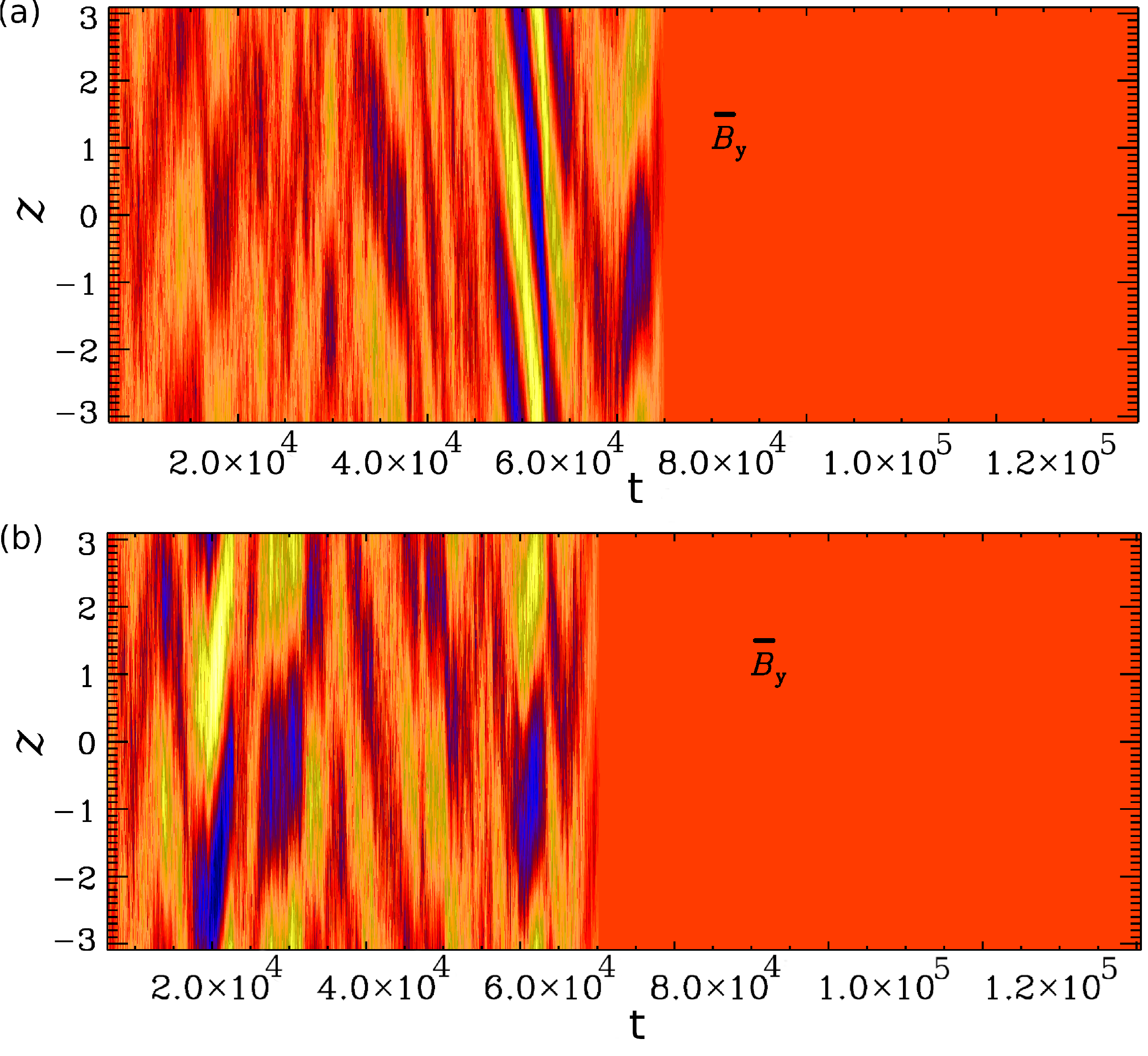}
\end{center}\caption[]{
Spacetime evolution of the horizontal averages of $B_y$ as a function of $z$ and time for (a) $\sigma = 0.198$ and (b) $\sigma = 0.197.$
 }\label{fig:By-mean}\end{figure}

For lower values of $\sigma$, the hydromagnetic attractor loses stability and becomes a hydromagnetic chaotic saddle, leaving a chaotic transient in the region of phase space previously occupied by the attractor. Two of such long chaotic transients are exhibited in Fig. \ref{fig:transient-0198-0197} in the form of time series of $B_{rms}$, and in Fig. \ref{fig:By-mean} in the form of the spatiotemporal evolution of the $B_y$ component averaged in the horizontal plane ($B_x, B_y$) as a function of $z$ and time. 

\begin{figure}[th!]\begin{center}
\includegraphics[width=\columnwidth]{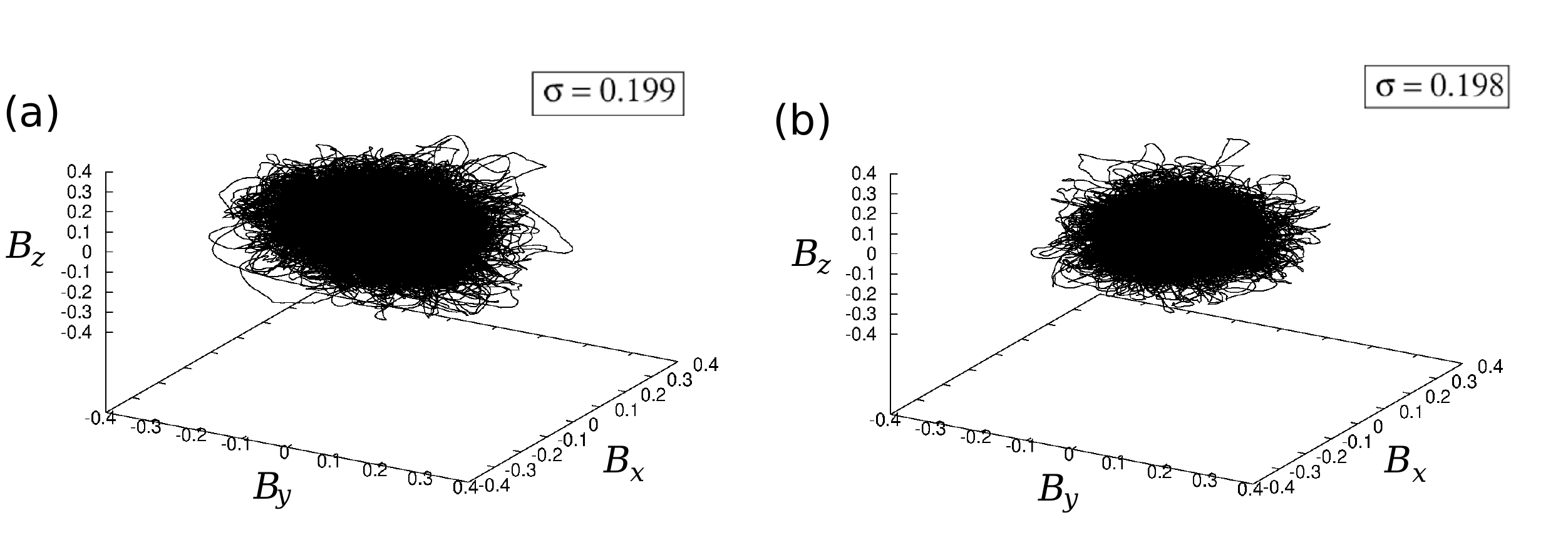}
\end{center}\caption[]{Trajectory of initial conditions on invariant sets in the $(B_x(x_0,y_0,z_0,t), B_y(x_0,y_0,z_0,t), B_z(x_0,y_0,z_0))$ space for the hydromagnetic attractor at $\sigma=0.199$ (a) and for the hydromagnetic chaotic saddle at $\sigma=0.198$ (b). 
The point $(x_0, y_0, z_0)$ is at the origin of the spatial domain.}\label{fig:cs}\end{figure}

The hydromagnetic chaotic attractor is shown in Fig. \ref{fig:cs}(a) for $\sigma=0.199$, where the chaotic trajectories represent the temporal variation of the components of the magnetic field vector $\textit{\textbf{B}}=(B_x(x_0,y_0,z_0,t), B_y(x_0,y_0,z_0,t), B_z(x_0,y_0,z_0,t))$ computed at the origin of the spatial domain $(x_0,y_0,z_0)=(0,0,0)$. A hydromagnetic chaotic saddle is shown in Fig. \ref{fig:cs}(b) for $\sigma=0.198$. It hints that both chaotic sets occupy approximately the same region in the phase space, but a small variation in the parameter $\sigma$ is sufficient to considerably reduce the size of the chaotic set, following the rapid decay of the magnetic energy in the upper branch of the bifurcation diagram in Fig. \ref{diagram-bifurcation}. 

\begin{figure}[th!]\begin{center}
\includegraphics[width=0.5\columnwidth]{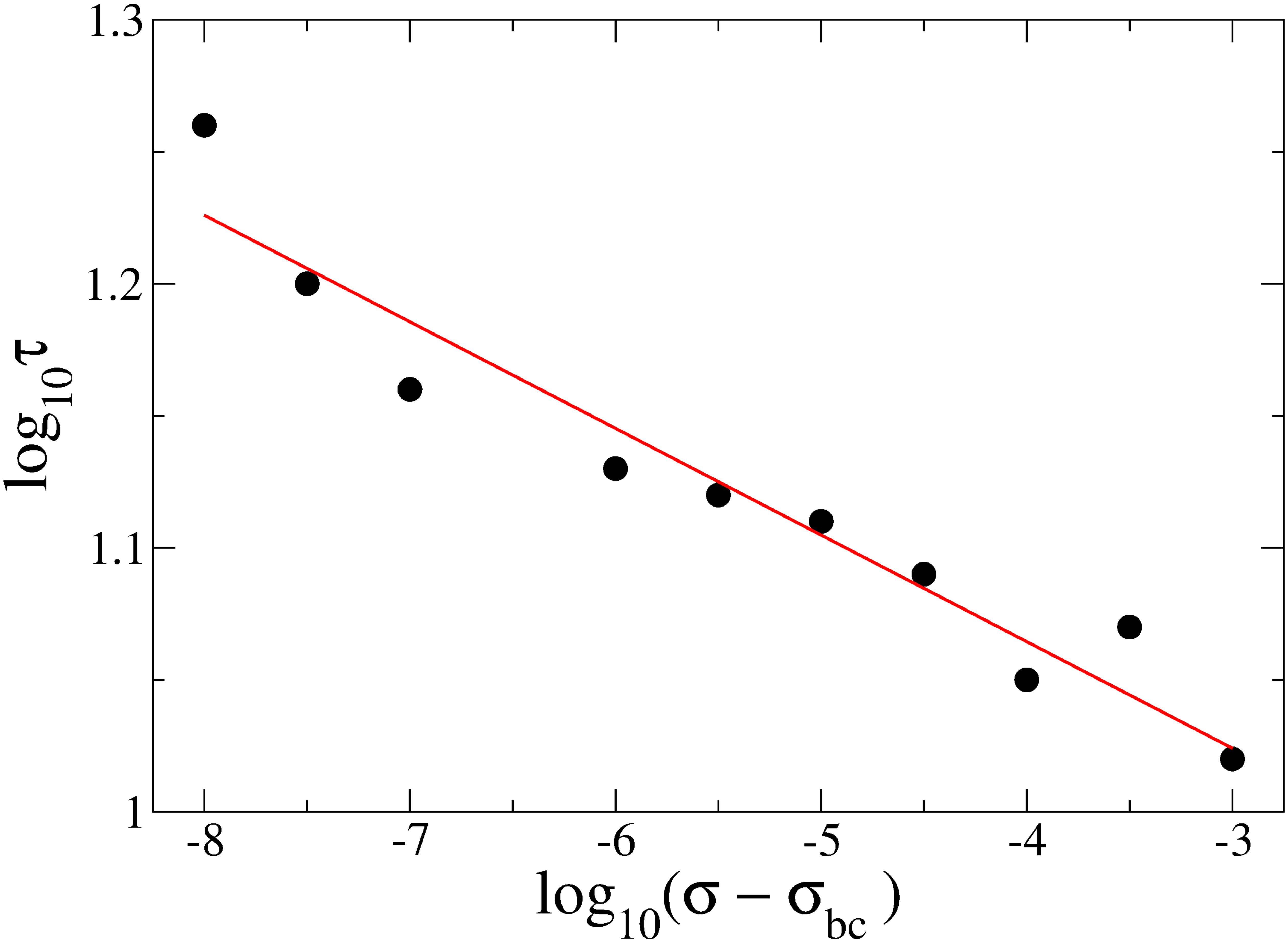}
\end{center}\caption[]{
Scaling law of the hydromagnetic chaotic transients as a function of the distance to the boundary crisis value $\sigma_{bc}$.
}\label{fig:scalinglaw}\end{figure}

As illustrated by Fig. \ref{fig:transient-0198-0197}, the average duration of the chaotic transients decreases with decreasing $\sigma$. This observation suggests that the destabilization of the hydromagnetic chaotic attractor at $\sigma=\sigma_{bc}\approx0.198$ is due to a boundary crisis, where a chaotic attractor collides with the boundary of its own basin of attraction, leading to the destruction of both the attractor and its boundary. A chaotic saddle is then left in place of the chaotic attractor and, as shown by Grebogi et al. \cite{Grebogi1987}, the average duration $\tau$ of the chaotic transients near a boundary crisis decays with the distance from the crisis parameter value $\sigma_{bc}$ following the law

\begin{equation}
\tau \sim (\sigma - \sigma_{bc})^{\gamma},
\end{equation}
for a negative $\gamma$. We computed $\tau$ for a set of values of $\sigma$ close to $\sigma_{bc}$ and obtained the results shown in Fig. \ref{fig:scalinglaw}, where the fitted line has slope $\gamma = -0.04$. The following procedure was adopted to produce this figure. First, a set of 100 initial conditions are selected from the hydromagnetic chaotic attractor at $\sigma=0.3 > \sigma_{bc}$; then, those initial conditions are used to generate transient chaotic time series for a given $\sigma<\sigma_{bc}$; the transient time for each initial condition is recorded when $B_{rms}$ is below a certain threshold ($B_{rms} < 0.01$) and the average transient time $\tau$ is computed from the 100 time series. This process is repeated for the 10 values of $\sigma < \sigma_{bc}$ shown in the figure. The fitted line was obtained by using linear regression.

Having established the existence of hysteresis in the current $\alpha^2$ dynamo model, we go back to the previously raised question of why strong intermittent bursts are not
observed in the transition to dynamo near $\sigma=0.21$. As mentioned before, the transition in the ABC dynamo is due to a nonhysteretic blowout bifurcation, where intermittency is present \cite{Sweet01}. Differently, the hysteresis in our model mean that there is a bistability region in the parameter space for $0.199 \leq \sigma \leq 0.213$ where two chaotic attractors coexist, the hydrodynamic and the hydromagnetic attractors.
Our results reveal that a hysteretic blowout bifurcation is responsible for the dynamo transition.
According to Ott and Sommerer \cite{Ott94}, in this type of bifurcation there is an attracting chaotic set in the (hydrodynamic) invariant manifold and this 
chaotic set loses stability as the system parameter $\sigma$ increases through a critical value $\sigma_c$ (as in Fig. \ref{diagram-bifurcation}).
For $\sigma < \sigma_c$ orbits not in the basin of the attractor on the invariant manifold go to some other (hydromagnetic) 
attractor off the manifold (as in Fig. \ref{histeresetime}). 
As $\sigma$ increases through $\sigma_c$ the attractor loses stability. 
For $\sigma$ slightly greater than $\sigma_c$, points started near the manifold can experience a chaotic transient in which their
orbits initially closely resemble those on the $\sigma < \sigma_c$
manifold attractor, but almost all points started near
the manifold eventually move off to the other attractor off the manifold (as in Fig. \ref{serie-021}(c)). 
Thus, on-off intermittency is not present in a hysteretic blowout bifurcation.

\begin{figure}[th!]\begin{center}
\includegraphics[width=0.7\columnwidth]{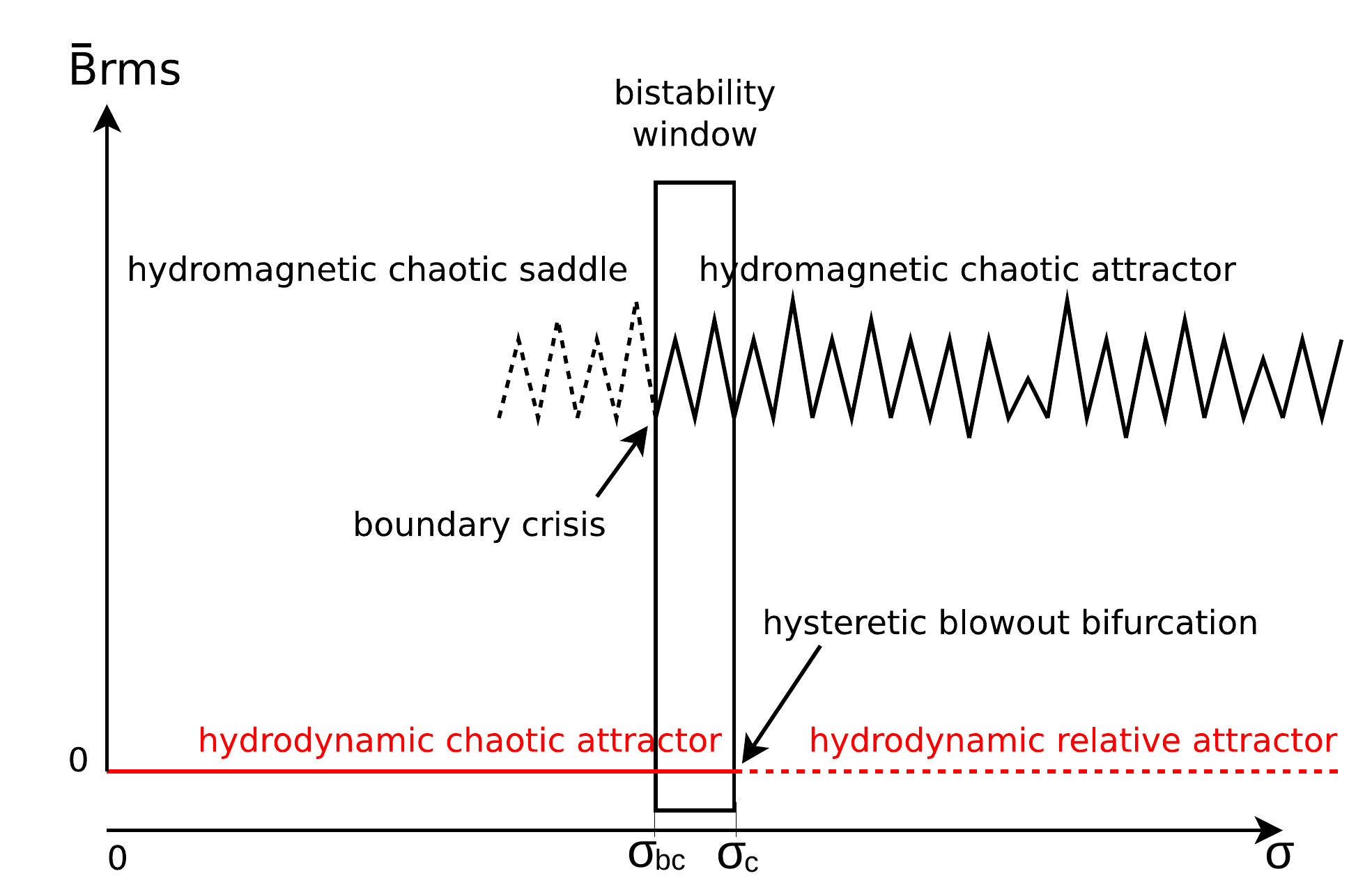}
\end{center}\caption[]{
Schematic bifurcation diagram of $\bar{B}_{rms}$ as a function of $\sigma$. Solid lines indicate attracting solutions and dashed lines indicate nonattracting solutions responsible for chaotic transients; the red lines represent hydrodynamic states and the broken black lines represent hydromagnetic chaotic states.
}\label{fig:Diagram1}\end{figure}

Our findings are summarized in Fig. \ref{fig:Diagram1}. The horizontal line at $\bar{B}_{rms}=0$ represents the hydrodynamic
manifold, which is attracting for $\sigma < \sigma_c$ (solid red line) and contains a chaotic attractor. For $\sigma > \sigma_c$, the hydrodynamic chaotic attractor
loses stability in a hysteretic blowout bifurcation and becomes a hydrodynamic ``relative attractor'' (dashed red line), where hydrodynamic chaotic 
transients are observed for initial perturbations near the hydrodynamic manifold. In the upper branch, the solid black line represents
the hydromagnetic chaotic attractor, which loses stability in a boundary crisis at $\sigma_{bc}$, giving rise to a hydromagnetic chaotic saddle (dashed black line), responsible for chaotic  transients involving the magnetic field. There is a bistability window between $\sigma_{bc}$ and $\sigma_c$ where the 
hydromagnetic chaotic attractor coexists with the hydrodynamic chaotic attractor.

\section{Conclusion}
\label{conclusion}

We have found a transition to dynamo in a hysteretic blowout bifurcation in direct numerical simulations of 
an MHD $\alpha^2$ dynamo model. Our transition differs from \cite{Sweet01} in that theirs is due to a non-hysteretic 
blowout bifurcation and it differs from \cite{KKB15} in that they adopted an $\alpha-\omega$ dynamo model. 
To our knowledge, this is the first time that a hysteretic blowout bifurcation is observed in an $\alpha^2$ dynamo model.
Previously, a nonhysteretic blowout bifurcation had been observed in a truncated mean field dynamo model \cite{covas}.
The hydromagnetic chaotic attractor in the upper hysteresis branch is destroyed in a boundary crisis.
Although we have not formally characterized the blowout bifurcation and the boundary crisis, the behavior
of the chaotic transients generated by our transitions strongly suggests that this is, indeed, the case. Thus, our work 
illustrates the importance of chaotic transients in revealing the complex dynamical transitions present in 
turbulent flows.

\ack
DNO and ELR acknowledge financial support from Brazilian agencies CAPES and CNPq. ELR acknowledges financial support from the Royal Society (UK). 
BBK acknowledges Department of Science and Technology (SERB/DST) for the financial support through the Ramanujan Fellowship (project no SB/S2/RJN-017/2018).
RC acknowledges support from the Funda\c c\~ao para a Ci\^encia e a Tecnologia (R\&D Unit SYSTEC: POCI-01-0145-FEDER-006933/SYSTEC funded by ERDF | COMPETE2020 | FCT/MEC | PT2020 extension to 2018, UID/EEA/00147/2020 and NORTE-01-0145-FEDER-000033 supported by ERDF | NORTE 2020) as well as the Russian Science Foundation (project no. 19-11-00258 carried out in the Federal Research Center ``Computer Science and Control'' of the Russian Academy of Sciences, Moscow, Russian Federation).

\section*{References}
\bibliography{references}

\end{document}